# Distinct magneto-Raman signatures of spin-flip phase transitions in CrI$_3$

Amber McCreary,[1] Thuc T. Mai,[1,2] Franz G. Utermohlen,[2] Jeffrey R. Simpson,[1,3] Kevin F. Garrity,[4] Xiaozhou Feng,[2] Dmitry Shcherbakov,[2] Yanglin Zhu,[5] Jin Hu,[6] Daniel Weber,[7,⊥] Kenji Watanabe,[8] Takashi Taniguchi,[8] Joshua E. Goldberger,[7] Zhiqiang Mao,[5] Chun Ning Lau,[2] Yuanming Lu,[2] Nandini Trivedi,[2] Rolando Valdés Aguilar,[2,§] and Angela R. Hight Walker[1,*]

[1]Nanoscale Device Characterization Division, Physical Measurement Laboratory, National Institute of Standards and Technology, Gaithersburg, Maryland 20899, USA
[2]Department of Physics, The Ohio State University, Columbus, OH 43210, USA
[3]Department of Physics, Astronomy, and Geosciences, Towson University, Towson, MD 21252, USA
[4]Materials Measurement Science Division, Material Measurement Laboratory, National Institute of Standards and Technology, Gaithersburg, Maryland 20899, USA
[5]Department of Physics, The Pennsylvania State University, University Park, PA 16802, USA
[6]Department of Physics, University of Arkansas, Fayetteville, AR 72701, USA
[7]Department of Chemistry and Biochemistry, The Ohio State University, Columbus, OH 43210, USA
[8]National Institute for Materials Science, 1-1 Namiki Tsukuba Ibaraki 305-0044, Japan
[⊥]Current address: Battery and Electrochemistry Laboratory, Institute of Nanotechnology, Karlsruhe Institute of Technology, 76344 Eggenstein-Leopoldshafen, Germany

*Email: angela.hightwalker@nist.gov

§Email: valdesaguilar.1@osu.edu

ABSTRACT: The discovery of 2-dimensional (2D) materials, such as $CrI_3$, that retain magnetic ordering at monolayer thickness has resulted in a surge of both pure and applied research in 2D magnetism. Here, we report a magneto-Raman spectroscopy study on multilayered $CrI_3$, focusing on two new features in the spectra that appear below the magnetic ordering temperature and were previously assigned to high frequency magnons. Instead, we conclude these modes are actually zone-folded phonons. We observe a striking evolution of the Raman spectra with increasing magnetic field applied perpendicular to the atomic layers in which clear, sudden changes in intensities of the modes are attributed to the interlayer ordering changing from antiferromagnetic to ferromagnetic at a critical magnetic field. Our work highlights the sensitivity of the Raman modes to weak interlayer spin ordering in $CrI_3$.

Magnetic van der Waals-bonded materials represent a rapidly growing research field,[1-6] where these materials provide a solid-state platform to study a variety of exciting physics and potential applications of magnetism in two dimensions, including proximity effects, control using strain and gating, spin fluctuations, magnetic excitations, spintronics, and possible quantum spin liquids.[7] One material of particular interest is chromium tri-iodide ($CrI_3$), a ferromagnet (FM) at bulk thicknesses below the Curie temperature ($T_c$) but with the remarkable property of layered antiferromagnetism (AFM) in thin multilayers.[2] While each individual layer is FM, the layers themselves are AFM coupled, and this effect persists for samples tens of layers thick. Furthermore, the interlayer spin arrangement in $CrI_3$ can be switched between AFM and FM by an electric field,[8-10] applied pressure,[11] and a magnetic field,[2,12-15] providing tunability in potential devices.

Raman spectroscopy is a powerful technique to study a variety of phenomena in 2D quantum materials, including effects of strain,[16] electron-phonon coupling,[17] phase transitions,[18] spin-phonon coupling,[19] and magnetic excitations.[20-22] Additionally, the diffraction-limited spot size allows for the investigation of atomically thin samples and heterostructures using a non-contact probe. In this work, we collect temperature- and magnetic field ($B$)-dependent Raman spectra on a thin (~10 layers) $CrI_3$ single crystal encapsulated in hexagonal boron nitride (hBN). Interestingly, at low temperature, increasing the magnetic field results in dramatic changes in the Raman spectra, indicating a magnetic field-induced phase transition when the interlayer spin arrangement changes from AFM to FM. By calculating the phonon dispersion of $CrI_3$ in both the FM and AFM state, we conclude that the new modes appearing in the AFM state are zone-folded phonons. This work validates magneto-Raman spectroscopy as a sensitive technique to probe interlayer magnetic ordering in quantum materials.

A thin flake of $CrI_3$ (≈7 nm from atomic force microscopy, or ≈10 layers) was encapsulated between two 20 nm to 30 nm flakes of hBN using the dry transfer technique.[22,23] In the *ab* plane, the $Cr^{3+}$ atoms are arranged in a honeycomb lattice, where each Cr atom is bonded with six I atoms to form a distorted octahedron (see Figure 1a and Figure S1). At room temperature, bulk $CrI_3$ has a monoclinic structure (C2/m) but exhibits a crystallographic phase transition near 220 K to rhombohedral ($R\bar{3}$),[25] where the main difference between the two structures is the stacking of the layers (see Figure S1). The bulk $T_c$ is around 61 K,[25] with the spins aligned perpendicular to the *ab* plane. Surprisingly, the magnetic behavior of thinner samples is very different; while the spins still align perpendicular to the *ab* plane, the interlayer magnetic stacking is AFM, as demonstrated through a variety of experimental techniques.[2,8,9,12-15,26,27] It has been theoretically[28-31] and experimentally[27,31] suggested that atomically thin $CrI_3$ does not go through the crystallographic phase transition that the bulk does, but instead remains in the monoclinic structure at low temperatures, resulting in AFM interlayer stacking.

Raman spectra were collected with a triple grating spectrometer using an excitation laser wavelength of 632.8 nm and keeping the power below 150 μW (≈1 μm spot size) at the sample to avoid heating. The laser polarization of the incoming $\varepsilon_i$ light makes an angle φ with respect to the *b* axis (Figure 1a), and the scattered $\varepsilon_s$ light angle θ is changed from θ = 0° (parallel, *xx*) to θ = 90° (perpendicular, *xy*). Since the crystallographic *a* and *b* axes are not known in our sample, the angle φ is an arbitrary, yet constant, angle in our experiments. Figure 1b shows the Raman spectra at *T* = 5 K, in both the *xx* and *xy* configurations. We confirm the monoclinic symmetry in our ≈10 layer flake by resolving two peaks at 108 cm$^{-1}$ and 109 cm$^{-1}$ between *xx* (black, scaled by 0.5) and *xy* (red), unlike the doubly-degenerate peak seen in the rhombohedral structure.[32] Thus, we label the phonons using the irreducible representations of the 2/m point group, where only the $A_g$ and $B_g$ modes are Raman active.[33,34] The density functional theory (DFT) calculated atomic displacements associated with these modes are shown in the Supplemental Information.

Two new modes appear below $T_c$ in the *xy* configuration at 77 cm$^{-1}$ (9.5 meV) and 126 cm$^{-1}$ (15.6 meV), labeled $P_1$ and $P_2$ in Figure 1b, respectively. These modes were previously attributed to magnon excitations since they appear in the magnetically ordered state and have their largest intensity in *xy* (inset of Figure 1b), indicating $B_g$ symmetry.[35] Instead, the bulk magnon dispersion[36] at 5 K shows a low-energy magnon at Γ below 1 meV (8 cm$^{-1}$), similar to what was measured by recent FM resonance (FMR) experiments,[37] and magnons at the M point of the Brillouin zone at ~9 meV and 15 meV. Furthermore, a recent Raman study of magnon excitations in FePS$_3$ showed it is possible for magnons to be present in both *xx* and *xy* in quasi-2D van der Waals magnets.[20] Thus, the $B_g$ nature of $P_1$ and $P_2$ is not conclusive evidence that they are magnons.

We studied the effects of an applied magnetic field on $P_1$ and $P_2$, as detailed in Figure 2. Two spectral ranges from 65 cm$^{-1}$ to 90 cm$^{-1}$ ($P_1$, $A_g^1$) and 120 cm$^{-1}$ to 136 cm$^{-1}$ ($P_2$, $A_g^6$) are shown on different intensity scales (≈ a factor of 3 to 1, respectively) for clarity. At *B* = 0 T and in *xy*, $P_1$ and $P_2$ have strong intensities, whereas the two $A_g$ modes at slightly higher frequencies have minimum intensities as they are forbidden in the *xy* configuration. Increasing the magnetic field results in drastic changes in the Raman spectra, where $P_1$ and $P_2$ behave in the same fashion. Above ≈1.6 T, the intensities of $P_1$ and $P_2$ abruptly start to vanish and $A_g^1$ and $A_g^6$ begin to appear in *xy*. By *B* = 2 T, $P_1$ and $P_2$ are absent in all polarization configurations, while $A_g^1$ and $A_g^6$ are no longer forbidden in *xy*. No further changes occur above 2 T, and no hysteresis was observed when the field was lowered back to 0 T. It should be noted that $P_1$ and $P_2$ do not show frequency shifting with magnetic field, suggesting they are not one-magnon processes with spins perpendicular to *ab*.

Raman spectra were collected as the magnetic field was increased in finer steps up to 2 T. This is shown as a false-color map (Figure 3a) and spectra (Figure 3b) for the frequencies near $P_2$ and $A_g^6$, where six distinct magnetic field ranges are revealed. Upon close inspection, there is additional Raman scattering intensity, *i.e.,* spectral weight, present between $P_2$ and $A_g^6$ around 128 cm$^{-1}$, although it cannot be discerned if the spectral weight is attributed to a single or multiple mode(s). No changes to $P_2$, $A_g^6$, or the spectral weight in-between were observed in Range 1 from 0 T and 0.6 T. In Range 2 between 0.7 T and 0.8 T, the spectral weight between them $P_2$ and $A_g^6$ appears to shift in frequency and intensity. The spectra are again stable through Region 3 from 0.8 T to 1.4 T, after which striking changes are seen in Regions 4 and 5. In Region 4, $P_2$ starts to decrease in intensity, the intensity of $A_g^6$ stays relatively constant, and the spectral weight in-between shifts in frequency and increases in intensity. In Region 5, $P_2$ and the spectral weight in-between both decrease in intensity until they disappear, while $A_g^6$ grows in intensity until $B > 1.95$ T (Region 6), when the phase transition is finally complete. It should be noted that experimental uncertainty, including instrument drift and corrections for Faraday rotation in magneto-cryostat objective lenses, can lead to small changes in peak intensities (generally less than 5%) when comparing consecutively taken Raman spectra. The intensity changes being tracked in the field ranges in Figure 3b, however, are more significant than any changes seen in the Γ point phonons (see Figure S3) and are thus outside of experimental uncertainty. Moreover, frequency shifts, such as those observed in the spectral weight between $P_2$ and $A_g^6$ in Regions 2 and 4, are significantly more reliable than intensity changes, generally reproducible to within one CCD detector pixel (≈0.4 cm$^{-1}$ with HeNe excitation herein). For fitted peaks (*e.g.,* well-defined phonons), the Raman shift frequency is even more precise (better than 0.1 cm$^{-1}$).

Recent magneto-tunneling measurements of few-layered CrI$_3$ also observed large changes in the tunneling current at nearly the same magnetic field values where we observe dramatic changes in the Raman spectra, such as at 0.8 T and 2 T.[12-15,26] These changes were attributed to the spin-filtering effect when the magnetic field is strong enough to change the interlayer spin arrangement from AFM to FM, and the effect was observed even for 20 nm thick samples (our sample is ≈ 7 nm). The striking resemblance in the evolution with magnetic field observed with magneto-tunneling, including sharp changes between regions of stability, to those reported herein implies Raman spectroscopy is detecting the phase transition caused by layers flipping spins from AFM to FM stacking. The observation of the first jump at 0.8 T in a variety of thicknesses of CrI$_3$ in magneto-tunneling[12-15,26] suggests it is most likely due to the surface layers (adjacent to the hBN) flipping while the second, final jump is the flipping of the internal layers at ≈ 2 T. This spin-flip phase transition is supported by the observation of the lower energy FM magnon for $B > 6.5$ T (Figure S4), which matches previous results on bulk, FM CrI$_3$,[36,37] and the lack of change in the Raman spectra between 2 T and 9 T (Figure S5). The 10 L sample remains monoclinic at high magnetic fields (Figure S6), demonstrating that a change in crystal structure is not the source of the evolution of the Raman spectra. In particular, the spectral weight present between $P_2$ and $A_g^6$ is extremely sensitive to the spin-flipping, displaying strong frequency shifts and intensity variations where the spin flips occur. The fact that the spectral weight does not shift in magnetic field in Regions 1 and 3, which are regions of stability, suggests its presence does not involve a one-magnon process. It is possible that the spectral weight results from the excitation laser wavelength ($\lambda$ = 632.8 nm) being nearly resonant with the ligand-to-metal charge transfer transition,[38-41] but the weak Raman signal at off-resonance excitation laser wavelengths makes this difficult to verify.

The polar plots for $A_g^6$, which track the intensity of $A_g^6$ as the angle θ (*i.e.*, $\varepsilon_s$ relative to $\varepsilon_i$) is changed, at various magnetic fields provide further evidence of the magnetic phase transition. As seen in Figure 4a, the polar plot is rotated by approximately 35° at $B$ = 2 T when compared with $B$ = 0 T. This can be understood if we write the Raman tensor for the $A_g$ phonon under applied magnetic field ($B \perp ab = B_z$) as:

$$R_{A_g, B \perp ab} = R_{A_g, B=0} + B_z \cdot R_{A_g, B_z} \quad (1)$$

The point group of CrI$_3$ requires that the total Raman tensor $R_{A_g}$ is symmetric under two-fold rotation around the *b*-axis ($C_{2x}$). Since $B_z$ itself is antisymmetric under $C_{2x}$, then the form of $R_{A_g, B_z}$ is required to also be antisymmetric under $C_{2x}$:

$$R_{A_g, B \perp ab} = \begin{pmatrix} \alpha & 0 \\ 0 & \beta \end{pmatrix} + B_z \begin{pmatrix} 0 & \gamma \\ \delta & 0 \end{pmatrix} \quad (2)$$

Thus, for the $B$ = 0 T case (AFM state), the Raman tensor would be purely symmetric, and we expect to not detect any signal in *xy* polarization configurations. A magnetic field applied perpendicular to the *ab* plane (FM state), however, introduces off-diagonal tensor elements and breaks this expectation, causing the polar plot to rotate as seen in Figure 4a. Assuming there are minimal changes between the monoclinic ($C_{2h}$), 10 L CrI$_3$ and the hexagonal point group of the monolayer ($D_{3d}$), then $\beta \approx \alpha + \Delta$ and $\delta \approx -(\gamma + \Delta')$, where $\Delta$ and $\Delta'$ are small corrections.

$$R_{A_g, +B_z} = \begin{pmatrix} \alpha & B_z \gamma \\ -B_z(\gamma + \Delta') & \alpha + \Delta \end{pmatrix}, \quad R_{A_g, -B_z} = \begin{pmatrix} \alpha & -B_z \gamma \\ B_z(\gamma + \Delta') & \alpha + \Delta \end{pmatrix} \quad (3)$$

Writing the matrices as in Eq. (3) makes it clearer that when the magnetic field direction is flipped from $B_z$ to $-B_z$, the signs of the off-diagonal elements switch, resulting in the opposite rotation of the polar plot, exactly as we observed in Figure 4b.

Figure 4c shows the angle of maximum intensity of the polar plot in 4a as the magnetic field is swept through the phase transition, where Regions 1-6 from Figure 3 are marked with vertical dashed lines. While Figure 3 and magneto-tunneling results indicate that parts of the phase transition occur between Regions 2 and 4, there is no observable change in the maximum intensity of the polar plot in those field ranges. In Region 5, however, it increases to ≈10° for $B = $ 1.7 T and 1.8 T, and then finally to ≈35° for 1.8 T and above. The fact that the observed change in the angle of maximum intensity of the polar plots is not linear with magnetic field, but instead remains constant until Region 5 indicates that $B_z$ should be regarded as the magnetization of the system induced by an external magnetic field rather than the field itself. The lack of rotation of the polar plot between 0.7 T and 0.8 T reveals that not enough magnetization is induced in this field range to introduce the off-diagonal Raman tensors in Eq. (2).

The temperature dependence of the spin-flips was investigated by tracking the intensity of $P_2$ as a function of magnetic field for different temperatures between $T = 9$ K and 26 K, as detailed in Figure 4d. The intensity of $P_2$ is shown relative to the intensity of the combination peak $A_g^5/B_g^3$ at ≈ 115 cm$^{-1}$ at $B = 0$ T for each temperature, as this peak appears to remain constant with temperature and magnetic field (in the probed range). The spin-flip transition field, or the amount of magnetic field necessary to cause $P_2$ to disappear, decreases with increasing temperature. In addition, the distinct jumps from spin flips and flat plateaus observed in the intensity of $P_2$ smooth out for higher $T$. Further temperature dependence is analyzed in Figure S8. This behavior is consistent with a phase transition where there is a strong correlation between the temperature and magnetic field, which suggests that the magnetic field takes the transition temperature to zero. While this behavior is akin to a quantum phase transition, there is no evidence yet of any quantum critical behavior in this material, but it is a very interesting avenue for future investigation.

Finally, we investigated the directional dependence of the spin-flip phase transition by rotating the sample such that the applied magnetic field is parallel to the *ab* plane ($B \parallel ab$). Figure 4e shows $I(P_2)/I(A_g^5/B_g^3)$ for $B \parallel ab$ and at $T = 2$ K. Unlike in the case for $B \perp ab$, no jumps or plateaus are observed in the intensity of $P_2$. Instead, the intensity of $P_2$ continuously decreases with increasing applied field up to $B = 6$ T, after which the layers are stacked FM with the spins pointing in the *ab* plane. This observation matches nicely with measurements of the magneto-tunneling current when the field is in the *ab* plane.[12-15] Furthermore, no frequency shift of $P_2$ is observed as a function of magnetic field (Figure S10). We perform the same tensor analysis done for Eqs. (1) and (2) above, but this time for the magnetic field applied in the *ab* plane.

$$R_{A_g, B \parallel ab} = R_{A_g, B=0} + B_x \cdot R_{A_g, B_x} + B_y \cdot R_{A_g, B_y} \quad (4)$$

where $x$ ($y$) corresponds to *b*-axis (*a*-axis). Requiring that the total Raman tensor be symmetric under $C_{2x}$, then $R_{A_g, B \parallel ab}$ is written as:

$$R_{A_g, B \parallel ab} = R_{A_g, B=0} + B_x \begin{pmatrix} \alpha' & 0 \\ 0 & \beta' \end{pmatrix} + B_y \begin{pmatrix} 0 & \gamma' \\ \delta' & 0 \end{pmatrix} . \quad (5)$$

From Eq. (5), when the magnetic field is directed along the *b*-axis and strong enough to align the spins along the *b*-axis, we expect the Raman tensor to be symmetric and the polar plot of the intensity of $A_g^6$ as a function of θ not to rotate. However, when the magnetic field has a component pointed along the *a*-axis, off-diagonal tensor elements are introduced, and some rotation of the polar plot would be expected. To test these predictions, we rotated the sample while the magnetic field was applied in the *ab* plane such that the magnetic field was pointed along two different crystal orientations $\varphi_1$ and $\varphi_2 = \varphi_1 + 90°$. The polar of the intensity of $A_g^6$ at $B = 0$ T and 7 T (in the spin-polarized state) are shown in Figure 4f. Interestingly, for both orientations, in which one of them must contain magnetic field components along the *a*-axis, the polar plot of the intensity of $A_g^6$ does not show the same rotation that was observed for spins pointing perpendicular to the *ab* plane in the spin-polarized state. This indicates that the off-diagonal elements $\gamma'$ and $\delta'$ are negligible for magnetic fields pointing in the *ab* plane.

Our data implies $P_1$ and $P_2$, which appear (disappear) in the AFM (FM) state, are not one-magnon excitations, as their frequencies do not shift with applied magnetic field. After considering other theoretical models (see Supplementary Information), we conclude that $P_1$ and $P_2$ are actually zone-folded phonons due to a doubling of the AFM unit cell in the *c*-direction. This is illustrated in Figure 5a, where the opposite spins of consecutive layers in the AFM configuration lead to a unit cell that is twice as large in the *c*-direction as compared to the FM configuration.

Using DFT, we calculated the phonon dispersions for monoclinic, bulk CrI$_3$ in both the FM and AFM stacking configurations. The full phonon dispersions, Brillouin zone, and table of frequencies are given in Figure S12 and Table S2. Figure 5b shows only the Raman active phonons in the FM state (red, solid) and AFM state (blue, dotted) for frequencies near $P_1$ and $P_2$. In general, Raman spectroscopy is only sensitive to modes at the Γ point in the phonon dispersion due to conservation of momentum. Thus, phonons at the A point (in direction of $k_z$) in the FM state are not observed. However, the doubling of the unit cell in the AFM state leads to zone-folding, where the phonons at the A-point in the Brillouin zone in the FM state fold back onto Γ in the AFM state and can be observed in Raman spectroscopy. Two pairs of modes with similar frequencies to $P_1/A_g^1$ and $P_2/A_g^6$ have an easily resolvable frequency splitting at Γ (highlighted Figure 5b), with the zone-folded phonon ($P_1$ or $P_2$) between 2 cm$^{-1}$ and 4 cm$^{-1}$ lower in frequency than the original phonon ($A_g^1$ or $A_g^6$). Illustrations of these vibrations are shown in Figures 5c and 5d, where the two layers vibrate in-phase for $A_g^1$ and $A_g^6$ and out-of-phase for $P_1$ and $P_2$. In the bulk, $P_{12}$ would have $B_u$ symmetry and would thus be Raman silent but infrared active. However, the breaking of inversion symmetry for an even number of layers in the AFM configuration would allow $P_{12}$ to be Raman active with $B$ symmetry (only seen in *xy* configurations). These zone-folded phonons would not shift in magnetic field and would be Raman-active (forbidden) in the AFM (FM) state, aligning with the observed behaviors of $P_1$ and $P_2$.

Even though the doubling of the unit cell is purely magnetic in nature, the observation of very strong zone-folded phonons that have similar intensity as the observed Γ point phonons reveals the strong coupling between magnetism and the lattice in atomically thin CrI$_3$. Of note, the origins of $P_1$ and $P_2$ as zone-folded phonons predict they would disappear for a monolayer sample (no zone-folding) or a sample with an odd number of layers because the AFM state preserves inversion symmetry with an odd number of layers. Recent work[42,43] has shown $P_1$ and $P_2$ are not present in monolayers of CrI$_3$, yet data by Jin *et al.*[35] suggests the presence of $P_1$ and $P_2$ for odd layer thicknesses. One possibility is that the encapsulation of thin layers of CrI$_3$ in hBN breaks inversion symmetry naturally, leading to the presence of $P_1$ and $P_2$ in all thicknesses. Further studies as a function of layer thickness and encapsulation parameters are needed to elucidate the symmetry response of $P_1$ and $P_2$ for even vs. odd numbers of layers.

In conclusion, we utilized magneto-Raman spectroscopy to elucidate a magnetic phase transition in $CrI_3$ where the interlayer stacking changes from AFM to FM. Substantial changes in the Raman spectra are detected at specific magnetic field values due to spin-flips of layers to a FM state, indicating that Raman modes are extremely sensitive to this phase transition. Moreover, Raman scattering proves to be crucial to understanding the symmetry and frequency shifts of the modes. We conclude that the new modes $P_1$ and $P_2$ are not high frequency magnons as previously believed, but instead attribute the modes to zone-folded phonons using symmetry arguments, polarization-dependent Raman responses, and calculated phonon dispersions in the FM and AFM stacking configurations. This study paves the way for further use of magneto-Raman spectroscopy to investigate spin-flip phase transitions in 2D van der Waals magnets.

**METHODS**

Sample Preparation and Encapsulation: Bulk $CrI_3$ crystals were grown by a chemical vapor transport technique using stoichiometric mixtures of Cr and I in a sealed evacuated quartz tube, as mentioned in other references.[25,44] The phase of the obtained crystals were checked by x-ray diffraction. These crystals were then exfoliated onto $Si/SiO_2$ substrates in an Ar-filled glovebox having $O_2$ and $H_2O$ concentrations of < 0.1 ppm. $hBN/CrI_3/hBN$ heterostructures were fabricated using a dry transfer technique detailed elsewhere.[23,24] Specifically, PDMS (polydimethylsiloxane) was used as the polymer stamp.

Raman Spectroscopy: Raman spectra were measured with the 632.8601 nm excitation wavelength of a He-Ne laser in the 180° backscattering configuration using a triple grating Raman spectrometer (Horiba JY T64000[†], 1800 $mm^{-1}$ grating) coupled to a liquid nitrogen cooled CCD detector. Polarization was selected and controlled using ultra broadband polarizers and achromatic half wave plates. To perform temperature- and magnetic-field dependent Raman, the sample was placed into an attoDRY cryostat (Attocube Inc.[†]), where the sample holder is pumped to $\approx 1 \times 10^{-3}$ Pa ($\approx 7 \times 10^{-6}$ Torr), backfilled with helium gas, and zero-field cooled. The few-micrometer sized flake of $CrI_3$ encapsulated in hBN was studied by focusing the laser with a white light camera onto the sample with a low-temperature, magnetic field compatible objective (50×, N.A. 0.82) and xyz nano-positioners. Integration times were approximately 12 minutes and the laser power was kept below 150 µW to reduce local heating of the sample. Spectra with applied magnetic field were corrected for Faraday rotation in the objective using half wave plates external to the cryostat.

DFT Phonon Calculations: We performed DFT calculations[45,46] with the Quantum Espresso[47] code, using the GBRV ultrasoft pseudopotential set.[48,49] We used the vdw-df-ob86 exchange correlation functional,[50,51] which includes long range van der Waals interactions, for our main results. We also tested the PBEsol[52] functional, finding similar results. Phonon calculations were performed using a finite differences (frozen phonon) approach, using PHONONPY[53] to perform symmetry analysis and the cluster_spring[54] code to calculate phonon dispersions. A 6 x 6 x 4 k-point sampling was used for the ferromagnetic unit cell.

†Certain commercial equipment, instruments, or materials are identified in this manuscript in order to specify the experimental procedure adequately. Such identification is not intended to imply recommendation or endorsement by the National Institute of Standards and Technology, nor is it intended to imply that the materials or equipment are necessarily the best available for the purpose.

ACKNOWLEDGEMENTS
A.M., T.T.M., and A.R.H.W. would like to acknowledge the National Institute of Standards and Technology (NIST)/National Research Council Postdoctoral Research Associateship Program and NIST-STRS (Scientific and Technical Research and Services) for funding. Work at The Ohio State University was supported by the Center for Emergent Materials, an NSF MRSEC under grant DMR-1420451. D.S. is supported by NSF/DMR 1807928. The single crystal growth efforts at Penn State are supported by the US Department of Energy under grant DE-SC0019068. D.W. gratefully acknowledges the financial support by the German Science Foundation DFG Research Fellowship (WE6480/1). Growth of hexagonal boron nitride crystals was supported by the Elemental Strategy Initiative conducted by the MEXT, Japan and the CREST(JPMJCR15F3), JST.

AUTHOR CONTRIBUTIONS
A.M., T.T.M, J.R.S., R.V.A, and A.R.H.W conceived the project. Bulk $CrI_3$ crystals were grown by Y.Z, J.H., D.W, J.E.G, and Z.M. Hexagonal boron nitride crystals were supplied by K.W. and T.T. Heterostructures were fabricated by D.S. and C.N.L. Raman measurements and analysis were carried out by A.M., T.T.M., J.R.S., R.V.A. and A.R.H.W. Theoretical analysis, including modeling and numerical calculations, were carried out by F.U. and X.F., N.T., and Y.-M.L. Phonon DFT calculations were carried out by K.F.G. All authors contributed to the writing of the manuscript.

## DATA AVAILABILITY

The data that support the findings of this study are available from the corresponding author upon reasonable request.

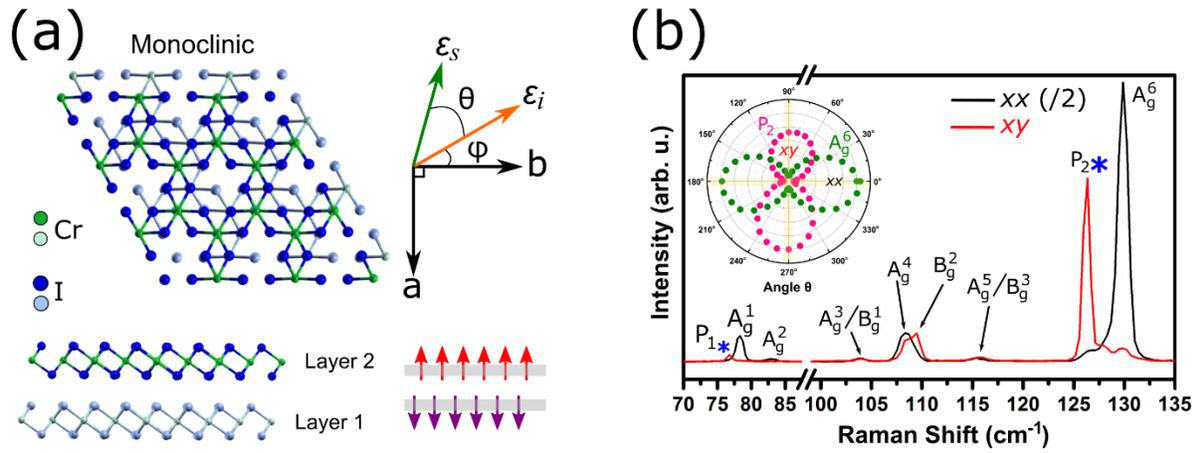

**Figure 1**: (a) Top and side views of two layers of $CrI_3$ with monoclinic structure and a schematic defining angles θ and φ with respect to the *a*- and *b*- crystal axes and the polarization vectors of the incoming $\varepsilon_i$ and scattered $\varepsilon_s$ light. The two $CrI_3$ layers are color-coded differently. (b) Raman spectra for both *xx* (black) and *xy* (red) configurations at $T = 5$ K and $B = 0$ T. Spectra in *xx* were divided by two for clarity. Inset shows intensity as a function of θ for $P_2$ and $A_g^6$ in a polar plot.

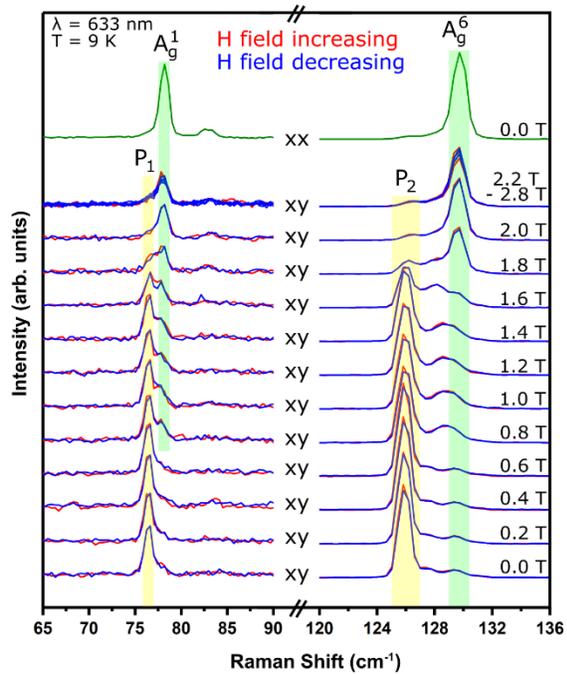

**Figure 2**: Magnetic field-dependent Raman spectra of CrI$_3$ from 0 T to 3 T ($B \perp ab$ plane) at $T = 9$ K where $P_1$ and $P_2$ disappear above 2.0 T while $A_g^1$ and $A_g^6$ appear despite being forbidden in *xy* configuration. A vertical offset was applied to spectra at different field values. *xx* spectrum at $B = 0$ T shown at top for comparison.

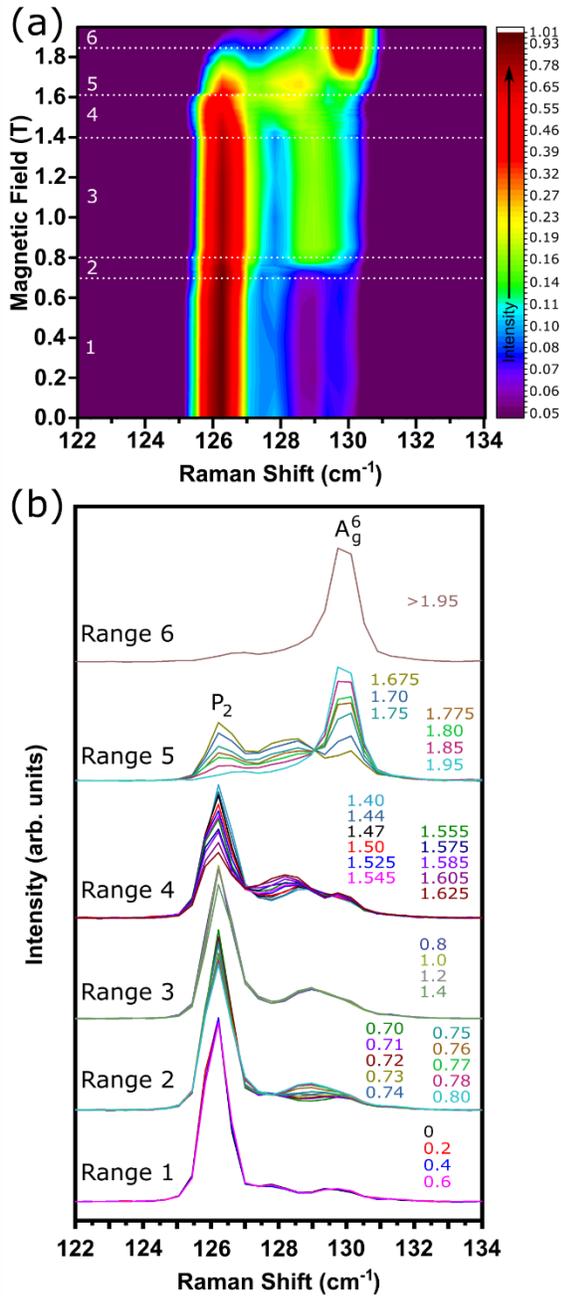

**Figure 3**: (a) False-color contour map of the Raman intensity (logarithmic) *vs.* magnetic field and shift frequency. (b) Raman spectra at $T = 9$ K, showing a detailed view of the 122 cm$^{-1}$ to 134 cm$^{-1}$ frequency range at different magnetic fields given by legend values (in Tesla). Six distinct field regions are identified below 2 T. A vertical offset was applied to spectra in the different field ranges for clarity.

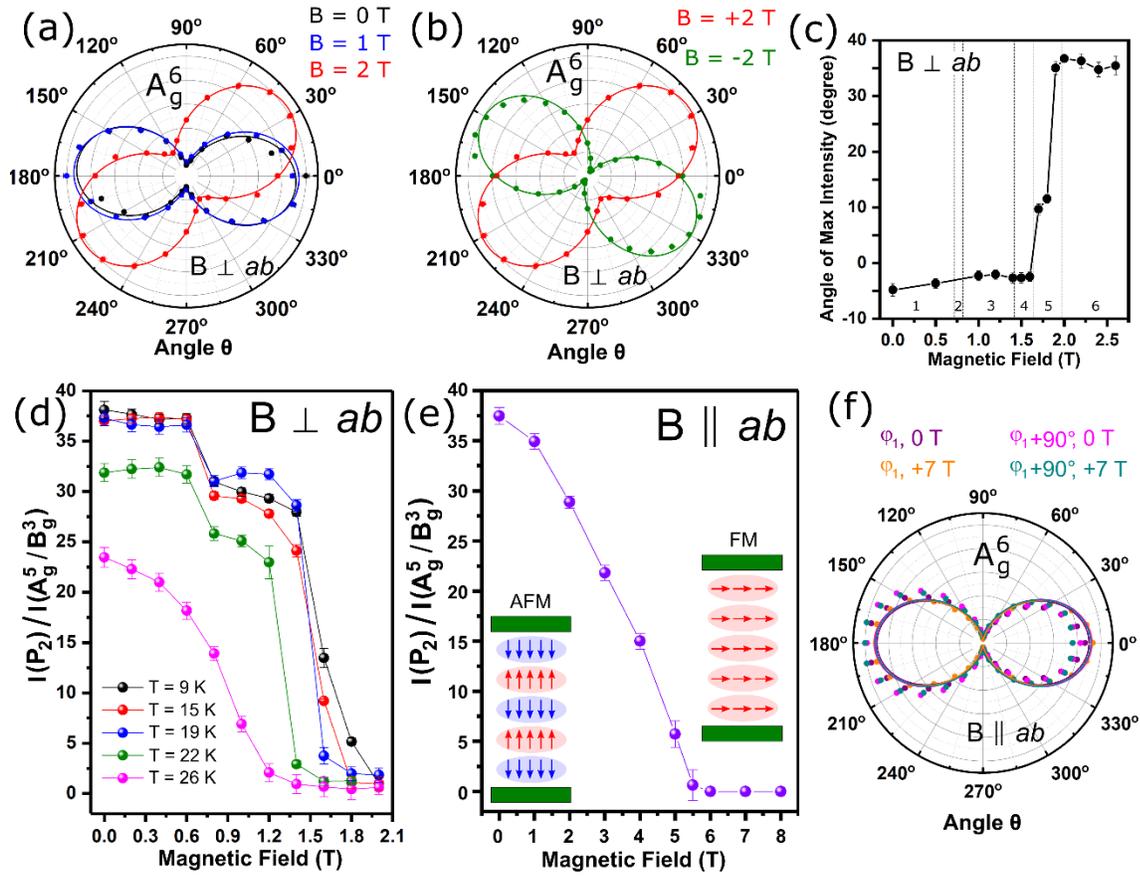

**Figure 4**: (a) Changes in polar intensity of $A_g^6$ (maximum intensity normalized to one), as a function of $\theta$ for (a) $B = 0$ T, 1 T, and 2 T ($B \perp ab$ plane). (b) Comparing the polar intensity of $A_g^6$ at positive and negative applied magnetic fields. (c) Angle of maximum intensity in polar plot in (a) as a function of magnetic field, where field ranges from Figure 3(b) are marked. (d) Intensity of $P_2$ relative to the intensity of the combined mode $A_g^5/B_g^3$, labeled $I(P_2)/I(A_g^5/B_g^3)$, as a function of applied magnetic field ($B \perp ab$ plane) at various sample temperatures. (e) $I(P_2)/I(A_g^5/B_g^3)$ as a function of applied magnetic field ($B \parallel ab$ plane) and (f) polar intensity of $A_g^6$ for a magnetic field (7 T) applied parallel to the $ab$ plane at two different crystal orientations. Sample temperature was 9 K (2 K) for $B \perp ab$ ($B \parallel ab$). Error bars represent standard errors from fitting function.

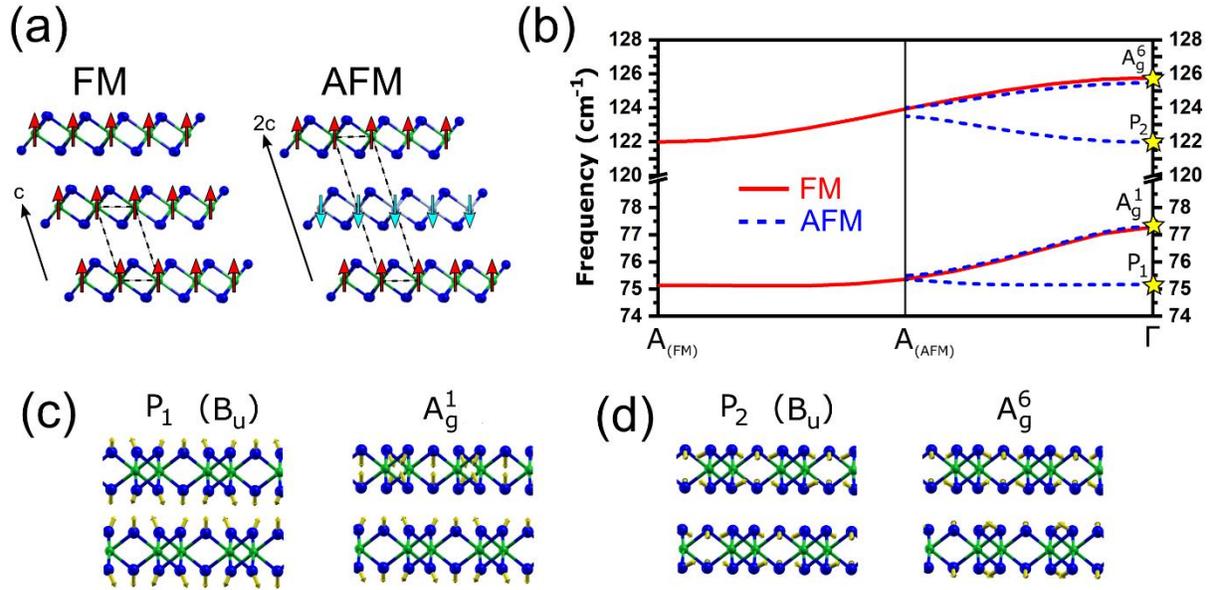

**Figure 5:** (a) Comparing the unit cell for the FM and AFM stacked structures, where the AFM unit cell is doubled in the *c*-direction. (b) Calculated phonon dispersion showing the Raman-active modes in the FM (red, solid) and AFM (blue, dashed) stacking in bulk $CrI_3$ in the monoclinic crystal structure. In the AFM case, the doubling of the unit cell along the *c*-direct in real space results in the $A_{(FM)}$-point folding into $\Gamma$, such that the number of modes is doubled compared to the FM case. The $\Gamma - A$ distance in the AFM case is half that of the FM case. Modes that correspond to $P_1$, $A_g^1$, $P_2$, and $A_g^6$ are highlighted and their atomic vibrations are shown in (c) and (d).

# SUPPORTING INFORMATION

# Distinct magneto-Raman signatures of spin-flip phase transitions in CrI$_3$

*Amber McCreary,[1] Thuc T. Mai,[1,2] Franz G. Utermohlen,[2] Jeffrey R. Simpson,[1,3] Kevin F. Garrity,[4] Xiaozhou Feng,[2] Dmitry Shcherbakov,[2] Yanglin Zhu,[5] Jin Hu,[6] Daniel Weber,[7,⊥] Kenji Watanabe,[8] Takashi Taniguchi,[8] Joshua E. Goldberger,[7] Zhiqiang Mao,[5] Chun Ning Lau,[2] Yuanming Lu,[2] Nandini Trivedi,[2] Rolando Valdés Aguilar,[2,§] and Angela R. Hight Walker[1,*]*

[1]Nanoscale Device Characterization Division, Physical Measurement Laboratory, National Institute of Standards and Technology, Gaithersburg, Maryland 20899, USA

[2]Department of Physics, The Ohio State University, Columbus, OH 43210, USA

[3]Department of Physics, Astronomy, and Geosciences, Towson University, Towson, MD 21252, USA

[4]Materials Measurement Science Division, Material Measurement Laboratory, National Institute of Standards and Technology, Gaithersburg, Maryland 20899, USA

[5]Department of Physics, The Pennsylvania State University, University Park, PA 16802, USA

[6]Department of Physics, University of Arkansas, Fayetteville, AR 72701, USA

[7]Department of Chemistry and Biochemistry, The Ohio State University, Columbus, OH 43210, USA

[8]National Institute for Materials Science, 1-1 Namiki Tsukuba Ibaraki 305-0044, Japan

[⊥]Current address: Battery and Electrochemistry Laboratory, Institute of Nanotechnology, Karlsruhe Institute of Technology, 76344 Eggenstein-Leopoldshafen, Germany

*Email: angela.hightwalker@nist.gov

§Email: valdesaguilar.1@osu.edu

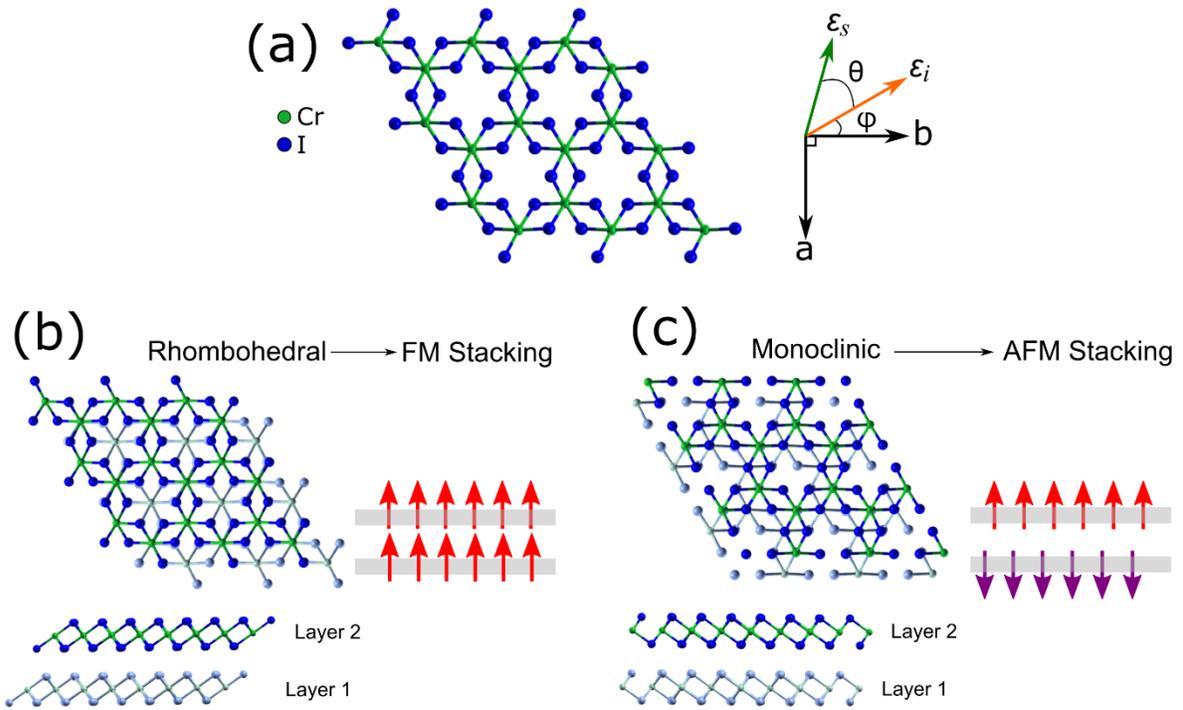

**Figure S1**: (a) Schematic showing top view of the crystal structure of a monolayer of $CrI_3$. (b,c) Comparing between the (b) rhombohedral and (c) monoclinic stacking of two layers of $CrI_3$, where the rhombohedral (monoclinic) results in ferromagnetic (antiferromagnetic) spin stacking.

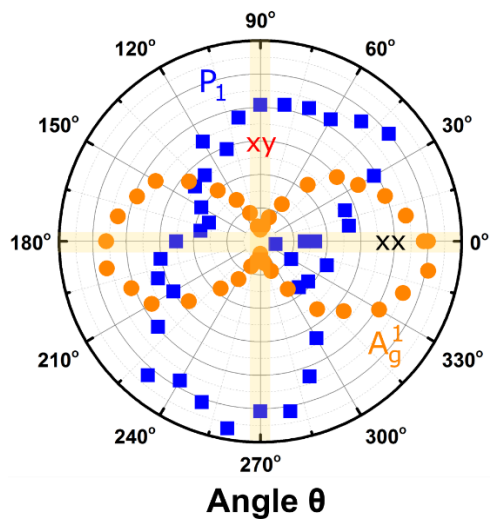

**Figure S2**: Polar intensity plot as a function of $\theta$ for $P_1$ mode and nearby $A_g^1$ phonon at T = 5 K and B = 0 T.

| Frequency (cm⁻¹) | Symmetry | Vibration |
|---|---|---|
| 51.09 | $B_g$ | |
| 52.15 | $A_g$ | |
| 77.00 | $A_g$ | |
| 87.92 | $B_g$ | |
| 100.99 | $A_g$ | |
| 101.22 | $B_g$ | |
| Continued… | | |

| Frequency (cm⁻¹) | Symmetry | Mode |
|---|---|---|
| 102.54 | $B_g$ | |
| 103.77 | $A_g$ | |
| 125.94 | $A_g$ | |
| 202.74 | $B_g$ | |
| 226.30 | $A_g$ | |
| 227.93 | $B_g$ | |

**Table S1**: DFT-calculated Raman-active phonons in monoclinic, bulk CrI$_3$.

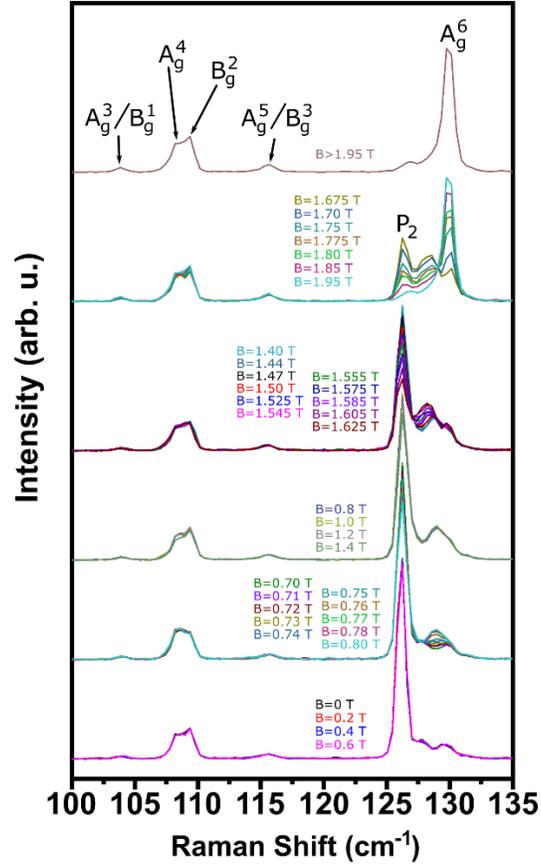

**Figure S3**: Raman spectra of 10 L CrI$_3$ as a function of magnetic field (applied perpendicular to the *ab* plane) at T = 9 K, showing a larger frequency range compared with Figure 3 of the main text where negligible changes are observed in the other phonon modes $A_g^3/B_g^1$, $A_g^4$, $B_g^2$, and $A_g^5/B_g^3$.

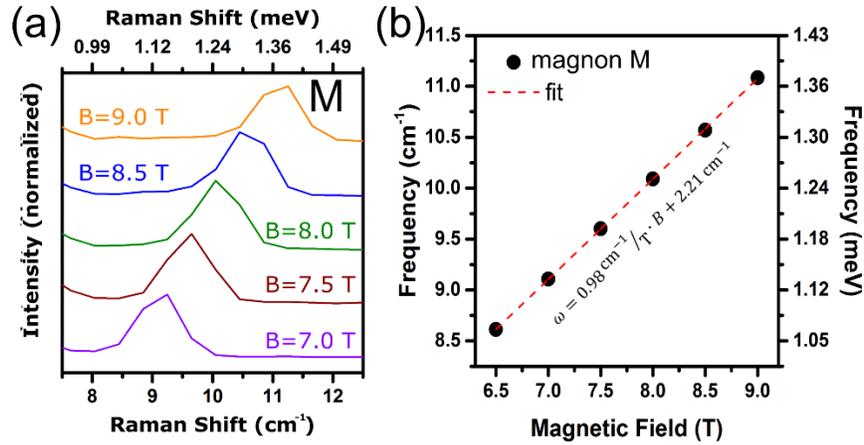

**Figure S4**: (a) Low-frequency Raman spectra at T = 5 K showing the true FM resonance (FMR), *i.e.* magnon, in CrI$_3$ that blueshifts with increasing magnetic field. At lower field values, we are unable to observe the FMR since it is below our spectrometer cutoff (~ 7 cm$^{-1}$). From the fit in (b), we extract $g \approx$ 2.08 ± 0.05 and $\omega_{B=0T}$ = 2.21 ± 0.3 cm$^{-1}$ (0.27 ± 0.04 meV).

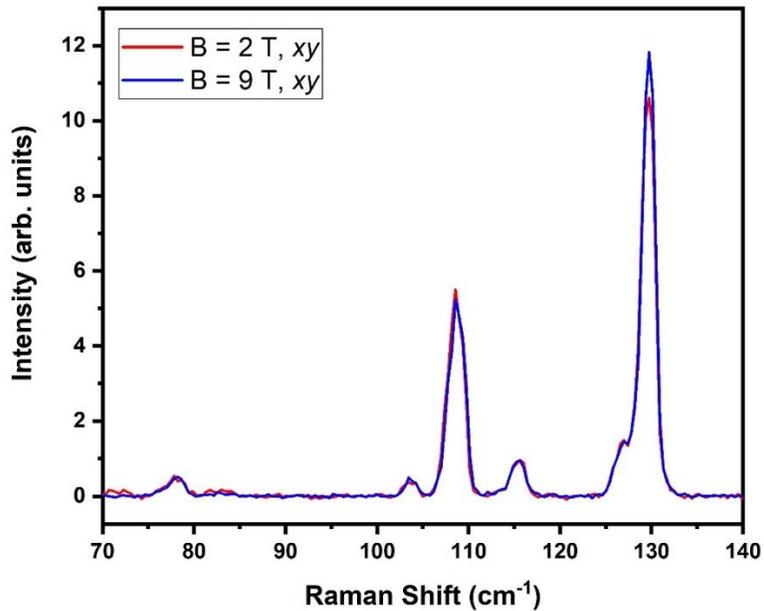

**Figure S5**: Comparing the Raman spectra (T = 9 K) of 10 L CrI$_3$ at B = 2 T and 9 T, where the magnetic field is applied perpendicular to the *ab* plane (*i.e.* parallel/antiparallel to the direction of the spins). Negligible changes are observed between the two spectra.

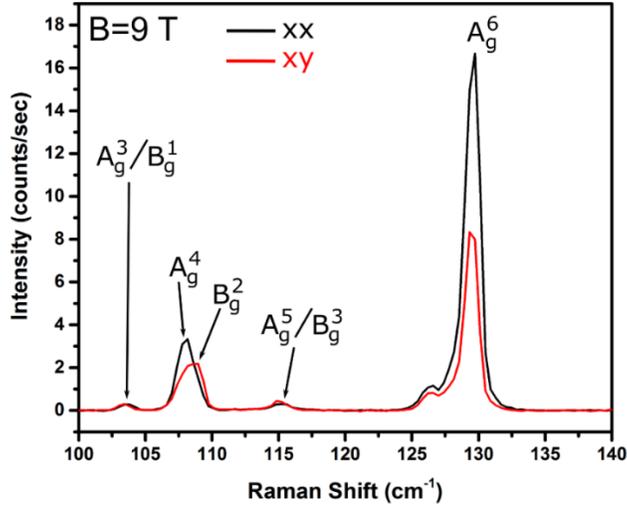

**Figure S6**: Comparing Raman spectra at B = 9 T (B ⊥ *ab*) and T = 9 K in the parallel (*xx*, black) and cross (*xy*, red) polarization configurations. Since we still observe a splitting in the mode at ≈108 cm$^{-1}$, as opposed to one degenerate mode, we deduce the thin CrI$_3$ is still in the monoclinic phase after the magnetic phase transition from AFM to FM interlayer stacking.

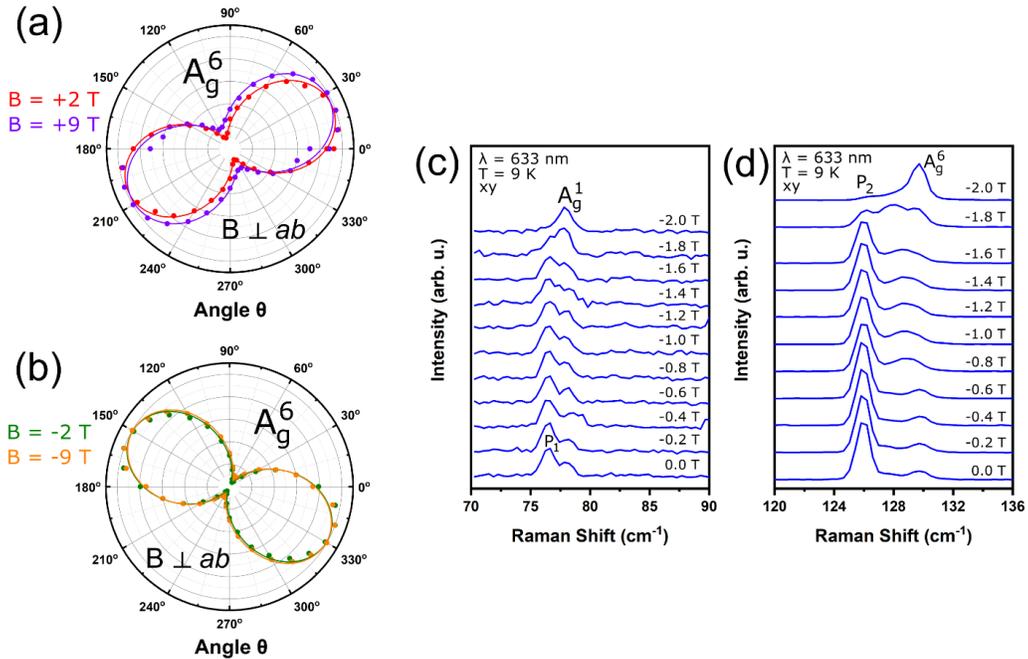

**Figure S7**: Comparing polar intensity plots for $A_g^6$ for (a) B = +2 T and +9 T and (b) B = −2 T and −9 T at 9 K, with the magnetic field applied perpendicular to the *ab* plane. Frequency range showing (c) $P_1$, $A_g^1$ and (d) $P_2$, $A_g^6$ as a function of applied negative magnetic field (B ⊥ *ab*), showing the same trend as applied positive magnetic fields.

**Temperature Dependence of Magnetic Phase Transition**

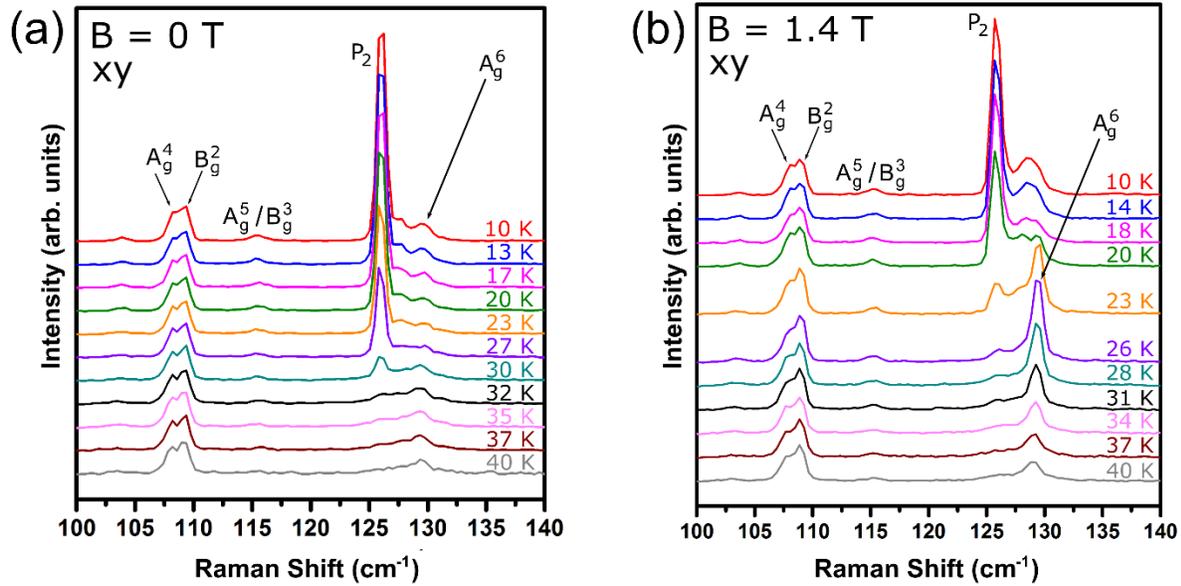

**Figure S8**: (a) Temperature dependence of the Raman spectra at B = 0 T in *xy* polarization configuration, where $P_2$ disappears above 32 K and the $A_g^6$ mode remains small. (b) Temperature dependence at B = 1.4 T, where increasing the temperature results in a similar evolution of the Raman spectra as further application of magnetic field, including the appearance/increased intensity of $A_g^6$, until it disappears again as the sample is warmed (B ⊥ *ab*).

      Figures S8a and S8b demonstrate that increasing temperature can have the same effect as increasing magnetic field to complete the phase transition in thin CrI$_3$. Figure S8a shows the evolution of the Raman spectra at B = 0 T, in *xy* configuration, as a function of temperature. As the temperature of the CrI$_3$ is increased, the intensity of $P_2$ decreases until it is unobservable above T = 35 K. In this case, with no magnetic field applied, the intensity of $A_g^6$ remains small, as expected. On the contrary, when a magnetic field slightly below $B_c$ is applied, such as B = 1.4 T in Figure S8b, increasing the temperature can be used in the same fashion as increasing the magnetic field to push CrI$_3$ through the phase transition, as was shown in magneto-resistance as well.[1] In this case, increasing the temperature by ≈16 K is enough to drive the phase transition that would have required an extra ≈0.6 T of magnetic field. With the magnetic field applied, the intensity of $A_g^6$ increases with increasing temperature, but as the temperature increases further towards the Curie temperature, $A_g^6$ starts to disappear again, confirming that the phase transition we are observing is magnetic in nature.

*In Figure S8b, the sample was zero field-cooled to T = 10 K, where a magnetic field of 1.4 T was applied perpendicular to the *ab* plane. Keeping the field at 1.4 T, the temperature was then increased to the values shown in the graph.

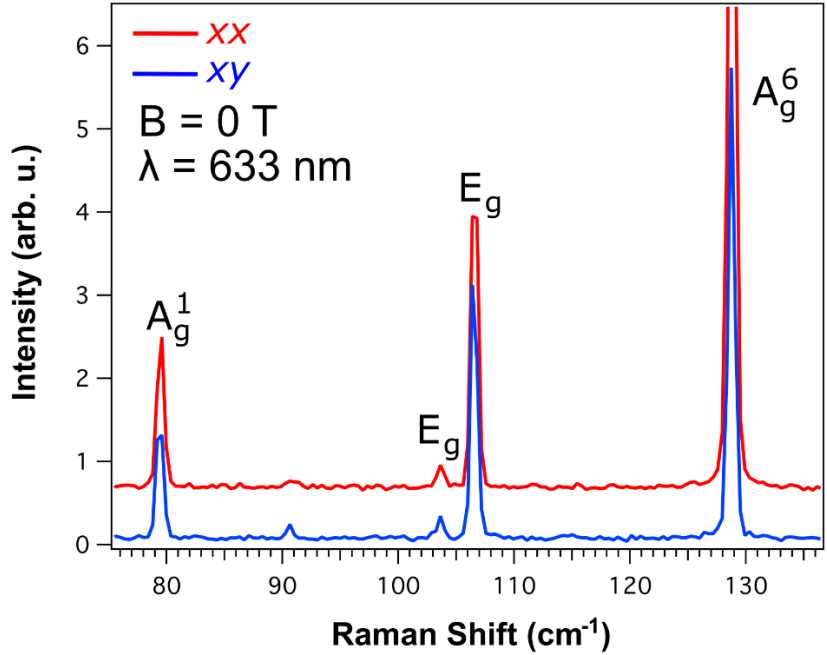

**Figure S9**: Raman spectra for bulk (≈1 mm thickness, not encapsulated in hBN) $CrI_3$ at T = 5 K (B ⊥ ab). The phonons are labeled using the rhombohedral point group $3\bar{R}$, since bulk $CrI_3$ does undergo a phase transition from monoclinic to rhombohedral at ≈200 K. Modes $P_1$ and $P_2$ were not observed, and in agreement with the rhombohedral structure, the mode at ≈107 cm$^{-1}$ is degenerate in *xx* and *xy* polarization configurations.

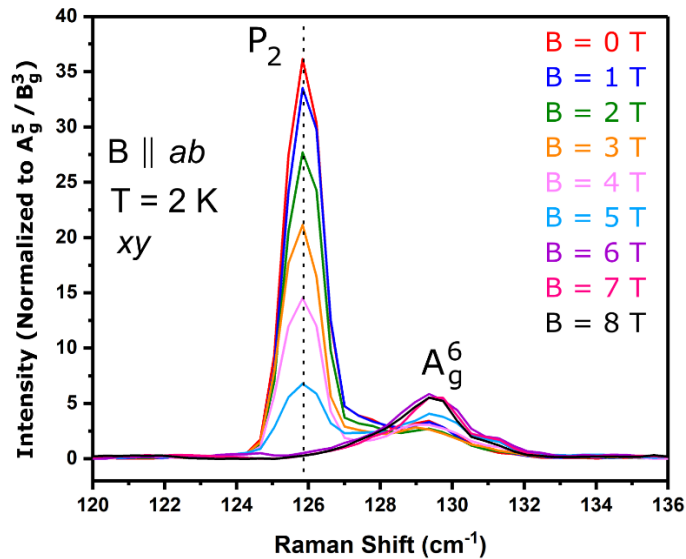

**Figure S10**: Evolution of $P_2$ and $A_g^6$ as a function of applied magnetic field when the field is applied parallel to the *ab* plane. Unlike the case of B ⊥ *ab*, a continuous decrease of $P_2$ as a function of field is observed. No frequency shift of $P_2$ is seen.

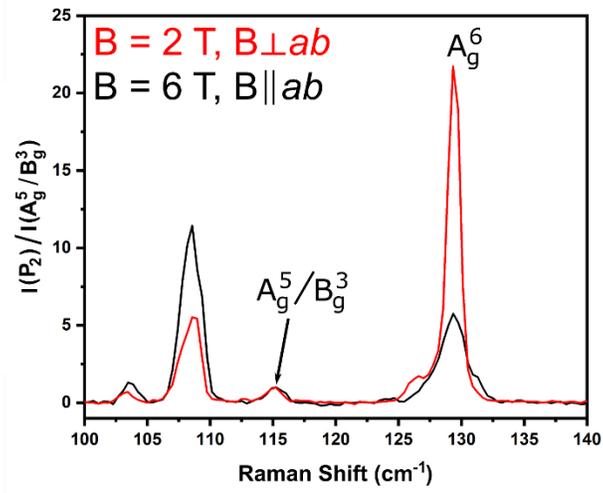

**Figure S11**: Comparing the Raman spectra (T = 9 K) above the magnetic-field driven spin-flip phase transition when the magnetic field is applied perpendicular (red) and parallel (black) to the *ab* plane.

# $P_1$ and $P_2$ as Zone-Folded Phonons:

**DFT Calculated Phonon Frequencies**

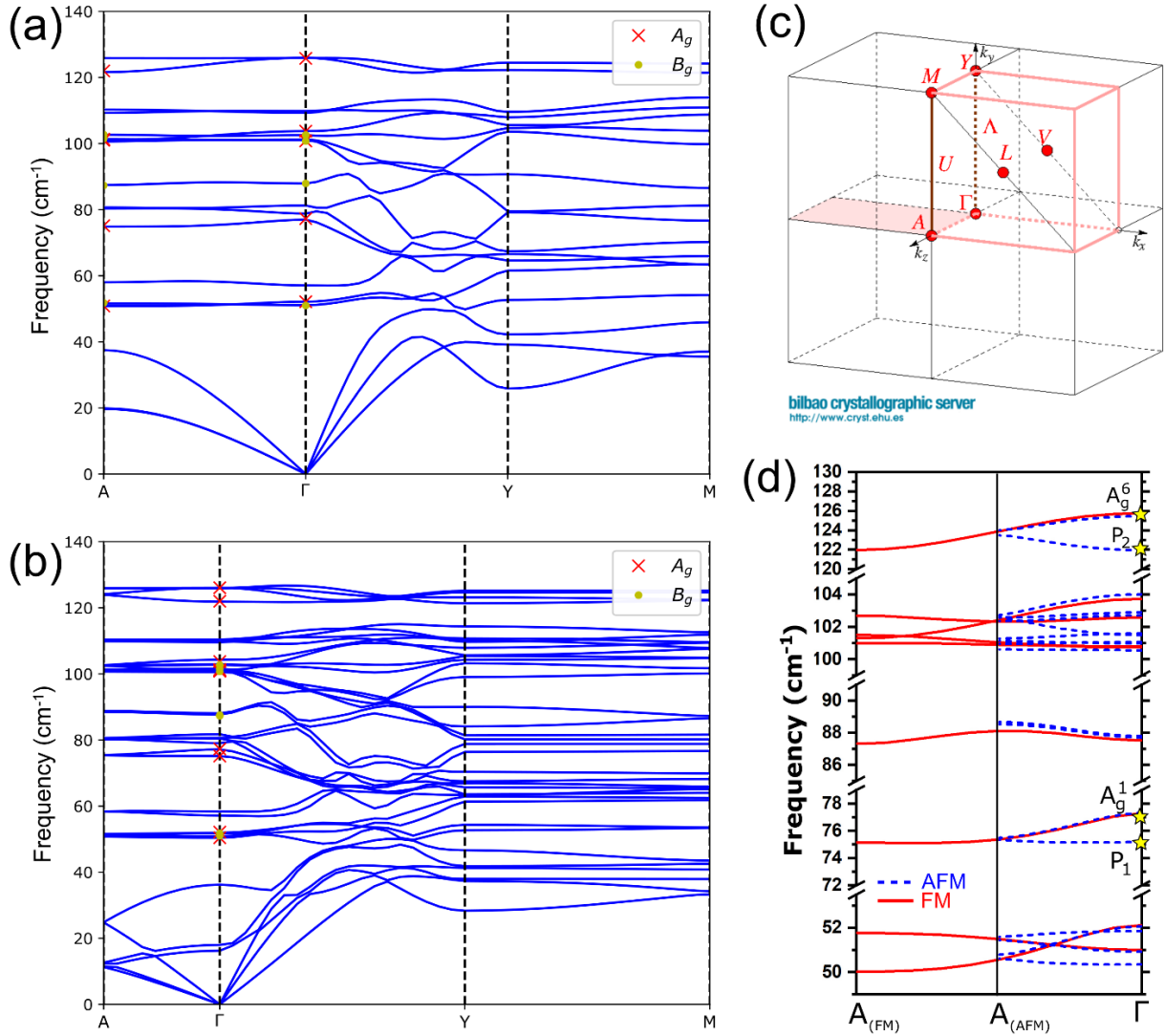

**Figure S12**: Calculated phonon dispersion of (a) FM and (b) AFM stacking for bulk CrI$_3$ in the monoclinic crystal structure. The relevant Brillouin zone from Bilbao crystallographic server[2-4] is shown in (c). In the AFM case, the unit cell doubles along the $c$-direction in real space, such that the A-point folds into Γ and the number of modes is doubled compared to the FM case. The symmetry of the Raman-active modes is marked on the phonon dispersion as $A_g$ or $B_g$ at the Γ point. It should be noted that a constant number of points was used between the high symmetry k-points for both the FM and AFM cases even though the size of the Brillouin zone in the Γ – A direction is half in the AFM vs. FM cases. Thus, it is not possible to visually compare the derivatives of the acoustic modes at Γ in order to get the speed of sound. (d) Comparing the calculated phonon dispersion showing the Raman-active modes between 45 cm$^{-1}$ and 130 cm$^{-1}$ in the FM (red, solid) and AFM (blue, dashed) stacking configurations. The Γ – A distance in the AFM case is half that of the FM case.

| Ferromagnetic (FM) | | Antiferromagnetic (AFM) | | |
|---|---|---|---|---|
| Frequency | Q-Point | Frequency | Q-Point | Character |
| 50.02 | A | 50.35 | Γ | oo-phase |
| 51.00 | Γ | 50.93 | Γ | in-phase |
| 51.77 | A | 51.85 | Γ | oo-phase |
| 52.12 | Γ | 52.07 | Γ | in-phase |
| 75.15 | A | 75.17 | Γ | oo-phase |
| 77.28 | Γ | 77.32 | Γ | in-phase |
| 87.31 | A | 87.70 | Γ | oo-phase |
| 87.53 | Γ | 87.80 | Γ | in-phase |
| 100.80 | Γ | 100.57 | Γ | in-phase |
| 100.73 | Γ | 100.98 | Γ | in-phase |
| 100.98 | A | 101.07 | Γ | oo-phase |
| 101.32 | A | 101.49 | Γ | oo-phase |
| 101.51 | A | 101.60 | Γ | oo-phase |
| 102.59 | Γ | 102.73 | Γ | mix-phase |
| 102.70 | A | 102.90 | Γ | mix-phase |
| 103.74 | Γ | 104.02 | Γ | in-phase |
| 121.98 | A | 121.96 | Γ | oo-phase |
| 125.77 | Γ | 125.47 | Γ | in-phase |
| 202.06 | Γ | 202.55 | Γ | in-phase |
| 201.87 | A | 202.67 | Γ | oo-phase |
| 225.54 | A | 226.31 | Γ | oo-phase |
| 226.36 | Γ | 226.70 | Γ | in-phase |
| 226.79 | A | 227.14 | Γ | oo-phase |
| 226.87 | Γ | 227.47 | Γ | in-phase |

**Table S2**: DFT-calculated phonon frequency and Q-points for the Raman-active phonons in monoclinically-stacked, bulk $CrI_3$ for ferromagnetic (FM) and antiferromagnetic (AFM) interlayer exchange coupling. For FM exchange, we list the phonon frequencies and symmetries at the A-point in the Brillouin zone as well, which is along the *c*-direction in real space. In the AFM state, the unit cell doubles and the A-point is folded into Γ and thus can become Raman-active. In general, the splitting between the mode originally at Γ and the mode zone-folded from the A-point are very small and most likely not resolvable. However, for the pairs of modes highlighted in red, the splitting is significant, with the mode from A at a lower frequency than the one from Γ.

| Frequency (cm⁻¹) | Q-Point in FM | Symmetry | Vibration |
|---|---|---|---|
| 75.2 | A | $B_u$ | |
| 77.3 | Γ | $A_g$ | |
| 122.0 | A | $B_u$ | |
| 125.8 | Γ | $A_g$ | |

**Figure S13**: DFT-calculated phonon vibrations for $A_g^1$ and $A_g^6$ as well as the zone-folded phonons at lower frequency. Eigenvectors are calculated for a bulk system, but are pictured as a bilayer system. Looking at the vibrations for a bilayer system, the two zone-folded vibrations would have $B_u$ symmetry with the inversion center in-between the two layers. When magnetic ordering is considered, however, where the two layers have AFM stacking, there is no longer inversion symmetry and the modes will be of *B* symmetry, and thus can be Raman active.

# $P_1$ and $P_2$ as Bound Two-Magnon Excitations:

We considered the possibility that $P_1$ and $P_2$ could be intralayer bound two-magnon excitations with a total spin of zero, given that they do not shift with $B$ field. A two-excitation consists of one magnon propagating in the $+k$ direction and other magnon propagating in the $-k$, such that $+k - k = 0$ and conservation of momentum is conserved, allowing these excitiations to be seen in Raman spectroscopy. If the bound two-magnon excitation had a total spin of zero, then the excitation would not be expected to shift in magnetic field, as was experimentally observed for $P_{1,2}$. In ferromagnetic systems, two-magnon excitations are expected to naturally arise from bond-dependent interactions like the Kitaev interaction,[5] which a recent experiment[6] suggests is the dominant interaction of CrI$_3$. We can see how the Kitaev interaction can produce two-magnon excitations by expressing the Kitaev interaction in the Holstein-Primakoff representation, which yields terms of the form $a_k^\dagger a_{-k}^\dagger$, corresponding to the creation of a pair of magnons with opposite momenta.

We performed an exact diagonalization calculation of the two-magnon density of states (DOS) and Raman intensity at zero temperature for a spin-3/2 system of 6-sites (a single honeycomb plaquette) described by the $JK\Gamma$ Hamiltonian

$$\hat{H}_{JK\Gamma} = \sum_{\langle r,r' \rangle \in \alpha\beta(\gamma)} \left[ J\bm{S}_r \cdot \bm{S}_{r'} + K S_r^\gamma S_{r'}^\gamma + \Gamma \left( S_r^\alpha S_{r'}^\beta + S_r^\beta S_{r'}^\alpha \right) \right]$$

which is the most general nearest-neighbor spin-spin interaction Hamiltonian allowed by the symmetries of a CrI$_3$ monolayer, where $\gamma \in \{x, y, z\}$ labels the bond type through which the neighboring Cr ions at $\bm{r}, \bm{r'}$ are interacting, $\alpha, \beta$ label the other two bond types, $J$ is the Heisenberg coupling, $K$ is the Kitaev coupling, and $\Gamma$ is the symmetric off-diagonal coupling. Using the values of the coupling constants obtained by Lee et al.,[25] namely $J = -0.212$ meV, $K = -5.190$ meV, and $\Gamma = -0.068$ meV, we found that the entire two-magnon DOS and Raman spectrum (see Figure S12) shift with applied $B$ field, thereby ruling out this potential mechanism for $P_1$ and $P_2$.

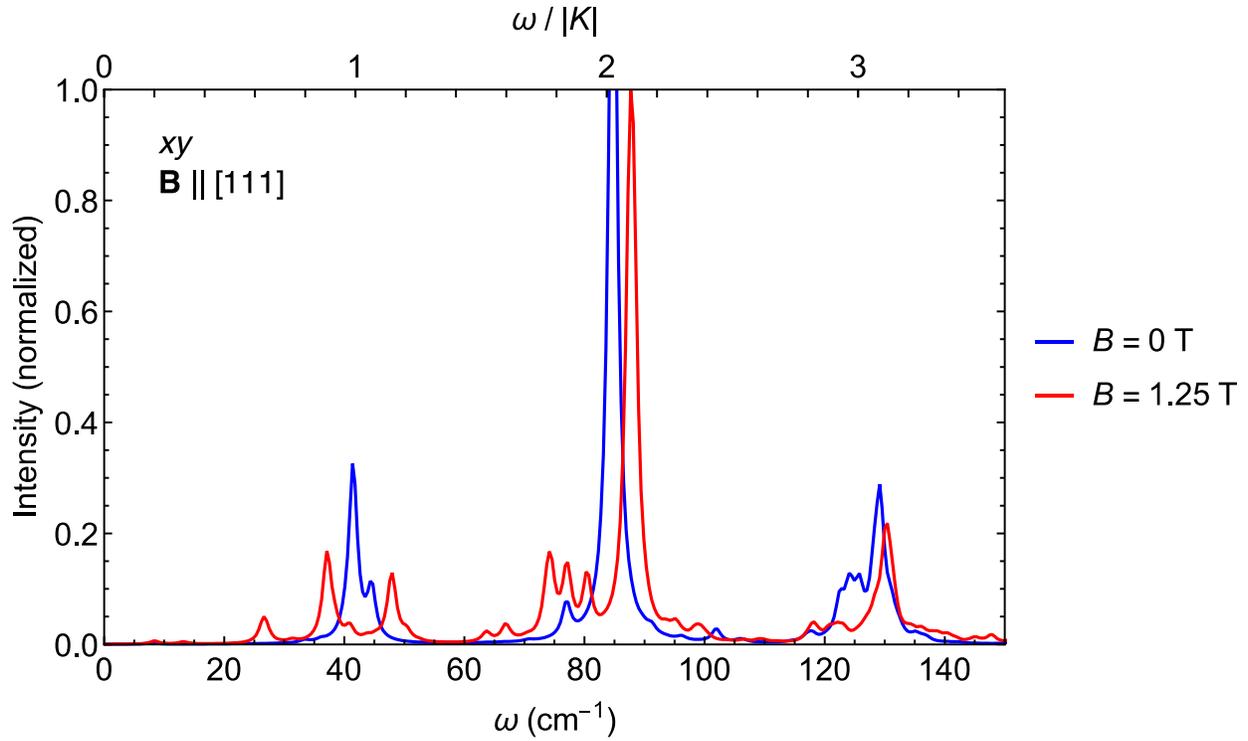

**Figure S14**: Exact diagonalization calculation of the zero-temperature Raman spectra in the cross (*xy*) polarization configuration for a spin-3/2 system of 6-sites (a single honeycomb plaquette) described by the *JK*Γ Hamiltonian and using the values of the coupling constants obtained by Lee *et al*.[25] This calculation serves as an estimate of the Raman behavior of monolayer $CrI_3$. When a *B* field is applied out of the plane, the entire Raman spectrum shifts.

## $P_1$ and $P_2$ as Bound Magnon-Phonon Pairs:

Another theoretical model we considered was that if the spins in CrI$_3$ were oriented in-plane and antiferromagnetically stacked, as has been reported in CrCl$_3$,[7] $P_1$ and $P_2$ could correspond to the bound state of a phonon and a magnon, where the phonon would be at a slightly higher frequency ($A_g^1$ for $P_1$, and $A_g^6$ for $P_2$) and the magnon would be a soft magnon $\gamma_{(0,0,\pi),+}$. We considered this as an option because the magnon (and hence the phonon-magnon bound state) carries $B_g$ representation of the $2'/m$ magnetic group in our 10-layer sample and exhibits the highest intensity in cross polarization, as was observed for $P_{1,2}$. The associated magnetic group $2'/m$ generated by magnetic rotation $C'_{2x}$ and mirror $M_x$ for the spins oriented in the *ab* plane is also consistent with the angle-dependence of second harmonic generation (SHG) signals observed by Z. Sun *et al.*,[8] since mirror $M_x$ and 2-fold rotation $C_{2x}$ enforces similar constrains to nonlinear susceptibility with respect to in-plane polarization. When the perpendicular magnetic field $B_z$ exceeds the critical value of 2 T, the in-plane order $AFM_x$ could be destabilized and transitioned into out-of-plane ferromagnetism $FM_z$ at larger fields.

We use the isotropic Heisenberg interaction between nearest neighbors as a minimal model in order to extract the symmetry properties of the magnon modes. The acquired representation of magnon modes and selection rules from Raman scattering are expected to be universal, insensitive to the exact form of the Hamiltonian. However, the frequency of the magnon modes can be an artifact of such toy models and should not be compared directly with experiments.

We label the lattice sites of the layered honeycomb lattice by

$$\boldsymbol{R} = \boldsymbol{r} + z\vec{a}_3 \equiv (r_1, r_2, z, s), \quad s = A/B \text{ labeling two sublattices} \tag{1}$$

$$\boldsymbol{r} \equiv r_1 \vec{a}_1 + r_2 \vec{a}_2 + \vec{R}_s, \quad r_{1,2}, z \in \mathbb{Z} \tag{2}$$

where $\{\vec{a}_i | i = 1,2,3\}$ are the primitive lattice vectors, and $\boldsymbol{r}$ is the in-plane 2D coordinate.

The monoclinic space group $2/m$ is generated by the following point group symmetries:

$$(x, y, z) \xrightarrow{M_x} (-x, y, z) \tag{3}$$

$$(x, y, z) \xrightarrow{C_{2x}} (x, -y, -z) \tag{4}$$

and their combination is the inversion symmetry.

The following minimal model of spin-3/2's is considered:

$$\hat{H} = -J \sum_{\langle r,r'\rangle,z} S_{r,z} \cdot S_{r',z} + J_z \sum_{r,z} S_{r,z} \cdot S_{r,z+1} - B_z \sum_R S_R^z \tag{5}$$

where $J \gg J_z, B_z > 0$ are all positive parameters.

## I.    LARGE FIELD CASE: OUT-OF-PLANE FERROMAGNETISM

For ferromagnetism (FM) along $\hat{z}$-axis (coined $FM_z$), we use the following Holstein-Primakoff representation to derive the spin wave theory:

$$S_R^z = S - a_R^\dagger a_R, \qquad S_R^+ = S_R^x + i S_R^y = \sqrt{2S - a_R^\dagger a_R} \cdot a_R \tag{6}$$

In this $FM_z$ phase, the unbroken magnetic symmetries (magnetic point group $2'/m'$) are

$$M_x' \equiv M_x \cdot \mathrm{T}, \quad C_{2x}' \equiv C_{2x} \cdot \mathrm{T} \tag{7}$$

where T is the time-reversal operator. The magnon operator transforms under them as

$$a_R \xrightarrow{M_x'} -a_{M_x R}, \quad a_R \xrightarrow{C_{2x}'} -a_{C_{2x} R}, \tag{8}$$

The associated linear spin wave Hamiltonian reads

$$\hat{H}_{FM_z} = (3JS + B_z - 2J_z S) \sum_R a_R^\dagger a_R - JS \sum_{\langle r,r'\rangle,z} a_{r,z}^\dagger a_{r',z} + J_z S \sum_{r,z} a_{r,z}^\dagger a_{r,z+1} + h.c. \tag{9}$$

It's straightforward to obtain the two magnon branches with frequency:

$$\omega_{k,k_z,\pm} = JS(3 \mp |f_k|) + B_z - 2J_z S(1 - \cos k_z) \tag{10}$$

where $\mathbf{k}$ labels the in-plane 2D momentum. $f_{\mathbf{k}} = \sum_{j=1,2,3} e^{i\mathbf{k}\cdot\vec{\delta}_j}$ is the structure factor of honeycomb nearest neighbors as in graphene, satisfying $f_{\mathbf{k}=(0,0)} = 3$. The wavefunction of the two magnon modes are given by

$$\gamma_{\mathbf{k},\pm} \equiv (a_{\mathbf{k},A} \pm a_{\mathbf{k},B})/\sqrt{2} \tag{11}$$

It is straightforward to check how they transform under symmetries:

$$\gamma_{\mathbf{k},\pm} \xrightarrow{M_x'} \mp \gamma_{-M_x \mathbf{k},\pm}, \qquad \gamma_{\mathbf{k},\pm} \xrightarrow{C_{2x}'} -\gamma_{-C_{2x}\mathbf{k},\pm} \tag{12}$$

since $M_x$ switches two sublattices but not $C_{2x}$. For magnons at the zone center $\Gamma$, the soft $\gamma_{\Gamma,+}$ mode is odd under both symmetries and belongs to $B_g$ representation of point group $2/m$. On the other hand, the high frequency $\gamma_{\Gamma,-}$ mode is odd under inversion symmetry and hence not Raman active. This explains why experiments only observe the soft $B_g$ mode $\gamma_{\Gamma,+}$ mode above $B_z \geq 7$ T, but not the high frequency mode $\gamma_{\Gamma,-}$.

## II. SMALL FIELD CASE: IN-PLANE ANTIFERROMAGNETISM

As suggested by the observation of SHG in bilayer CrI$_3$,[7] multilayer CrI$_3$ is likely to exhibit antiferromagnetism (AFM) between two neighboring layers. In the limit of a small magnetic field $B_z < 2$ T, we consider an in-plane moment along the $\hat{x}$-axis as reported in CrCl$_3$.[23] The Holstein-Primakoff representation writes:

$$S_R^x = (-1)^z \cdot (S - a_R^\dagger a_R), \quad S_R^{(-1)^z} \equiv S_R^x + (-1)^z i S_R^z = \sqrt{2S - a_R^\dagger a_R} \cdot a_R \tag{13}$$

For an even number of layers $L_z = 0$ mod 2, the magnetic point group is $2'/m$ generated by $C'_{2x}$ and $M_x$. The boson operator transforms as

$$a_R \xrightarrow{M_x} -a_{M_x R}, \quad a_R \xrightarrow{C'_{2x}} a_{C_{2x}R} \tag{14}$$

The linear spin wave Hamiltonian writes

$$\hat{H}_{AFM_x} = \sum_R (3JS + 2J_z S) a_R^\dagger a_R - JS \sum_{\langle r,r' \rangle, z} a_{r,z}^\dagger a_{r',z} + J_z S \sum_{r,z} a_{r,z}^\dagger a_{r,z+1}^\dagger + h.c.$$
$$- i B_z \sqrt{2S} \sum_R (-1)^z (a_R^\dagger - a_R) \tag{15}$$

Four branches of magnons are obtained with frequency:

$$\Omega_{k,k_z,\pm} = \sqrt{\omega_{k,\pm}(\omega_{k,\pm} + 4J_z S) + (2J_z S \sin k_z)^2} \tag{16}$$

where $\omega_{k,\pm} \equiv JS(3 \mp |f_k|)$ is the FM magnon dispersion within each 2D honeycomb plane. In particular, the magnon dispersion is not affected by the out-of-plane magnetic field $B_z$. It would, however, be affected by an in-plane magnetic field $B_{x,y}$, which is not consistent with our experimental observations for $P_1$ and $P_2$

There are two soft magnon modes, $\gamma_{(0,0,0),+}$ and $\gamma_{(0,0,\pi),+}$. Their symmetry characters are summarized in Table S3. They belong to $A_u$ and $B_g$ representations of group $2/m$, and the $B_g$ mode $\gamma_{(0,0,\pi),+}$ is the only Raman active branch of soft magnons.

If the number of layers is odd, the crystal symmetry $2/m$ is fully preserved. Under symmetry operations, the bosons transform as

$$a_R \xrightarrow{M_x} -a_{M_x R}, \quad a_R \xrightarrow{C_{2x}} a_{C_{2x} R} \tag{17}$$

The associated symmetry representations of magnons are summarized in Table S4. In this case, both soft magnons are Raman active and belong to $B_g$ representation.

| Modes | $M'_x$ | $C'_{2x}$ | Irrep. | Raman active? |
|---|---|---|---|---|
| $\gamma_{k_z=0,+}$ | - | + | $A_u$ | No |
| $\gamma_{\pi,+}$ | - | - | $B_g$ | Yes |
| $\gamma_{k_z=0,-}$ | + | + | $A_g$ | Yes |
| $\gamma_{\pi,-}$ | + | - | $B_u$ | No |

**Table S3**. Symmetry characters (magnetic point group $2'/m$) of magnons in the $AFM_x$ phase, in a thin film of CrI$_3$ with an even number of layers.

| Modes | $M_x$ | $C_{2x}$ | Irrep. | Raman active? |
|---|---|---|---|---|
| $\gamma_{k_z=0,+}$ | - | - | $B_g$ | Yes |
| $\gamma_{\pi,+}$ | - | - | $B_g$ | Yes |
| $\gamma_{k_z=0,-}$ | + | - | $B_u$ | No |
| $\gamma_{\pi,-}$ | + | - | $B_u$ | No |

**Table S4**. Symmetry characters (magnetic point group $2/m$) of magnons in the $AFM_x$ phase, in a thin film of CrI$_3$ with an odd number of layers.

### III. SMALL FIELD CASE: OUT-OF-PLANE ANTIFERROMAGNETISM

Finally, we consider out-of-plane AFM order $AFM_z$ as a comparison to $AFM_x$. The calculation is straightforward and one can show its magnon spectra as

$$\Omega_{k,k_z,\pm,\eta_z=\pm 1} = \left| B_z + \eta_z \sqrt{\omega_{k,\pm}(\omega_{k,\pm} + 4J_z S) + (2J_z S \sin k_z)^2} \right| \tag{18}$$

Clearly, all magnon frequencies shift with the applied magnetic field $B_z$, which is not consistent with our data for $P_1$ and $P_2$.

For comparison, we also list the representations of magnon modes at zone center $\Gamma$, for the case of even (Table S5) vs. odd (Table S6) layers in the thin film. They are the same as in the $AFM_x$ phase, although their magnetic point groups are different from the $AFM_x$ phase.

| Modes | $M'_x$ | $C_{2x}$ | Irrep. | Raman active? |
|---|---|---|---|---|
| $\gamma_{k_z=0,+}$ | - | + | $A_u$ | No |
| $\gamma_{\pi,+}$ | - | - | $B_g$ | Yes |
| $\gamma_{k_z=0,-}$ | + | + | $A_g$ | Yes |
| $\gamma_{\pi,-}$ | + | - | $B_u$ | No |

**Table S5**. Symmetry characters (magnetic point group $2/m'$) of magnons in the $AFM_z$ phase, in a thin film of CrI$_3$ with an even number of layers.

| Modes | $M'_x$ | $C'_{2x}$ | Irrep. | Raman active? |
|---|---|---|---|---|
| $\gamma_{k_z=0,+}$ | - | - | $B_g$ | Yes |
| $\gamma_{\pi,+}$ | - | - | $B_g$ | Yes |
| $\gamma_{k_z=0,-}$ | + | - | $B_u$ | No |
| $\gamma_{\pi,-}$ | + | - | $B_u$ | No |

**Table S6**. Symmetry characters (magnetic point group $2'/m'$) of magnons in the $AFM_z$ phase, in a thin film of CrI$_3$ with an odd number of layers.

In summary, if $P_{1,2}$ were due to the bound state of a magnon and a phonon where the spins were aligned in-plane, they would not be expected to shift in frequency under the application of a magnetic field perpendicular to the *ab* plane, which follows the observed behavior of $P_{1,2}$. They would, however, be expected to display a shift in frequency when the magnetic field was applied in the *ab* plane, but this is not consistent with what was observed in Figure S10. Thus, we have ruled out that $P_{1,2}$ are due to the bound state of a magnon and a phonon.